# A Comprehensive Study on Cyber-Attack Vectors in EV Traction Power Electronics

Siddhesh Pimpale

Lead Application Engineer, Dana Incorporated

| ARTICLE INFO | ABSTRACT |
| --- | --- |
| Received: 26 Feb 2023<br><br>Accepted: 14 Apr 2023 | Electric vehicles (EVs) have drastically changed the auto industry and developed a new era of technologies where power electronics play the leading role in traction management, energy conversion and vehicle control processes. Nevertheless, this is a digital transformation, and the cyber-attack surface area has increased considerably, to the point that EV traction power electronics are becoming vulnerable to various cybersecurity risks. This paper is able to provide its expertise on possible cyber-attack vectors which can attack important parts of the traction, powertrain, including things like inverters, motor controllers, and communicated systems within the embedded bits. Using the (STRIDE) threat modeling framework, the research outlines and groups the vulnerabilities of the architecture and runs some attack simulations, such as the Denial-of-Service (DoS), spoofing, firmware manipulation, and data injection. The experiments prove the fact that a slight interruption in the control signal, the sensed data may lead to the severe working implications, such as unstable sensor values of the torque, abnormal voltage shifts, and entire system freezes. These results highlight the high priority on the need of injective embedded intrusion preventive mechanisms and secure design of firmware in EV powertrain electronics. In this paper, the author makes his contribution to the general body of knowledge that underpins the links existing between cyber security practices and the peculiar needs of automotive power electronics. It provides feasible suggestions to the manufactures of the vehicle, designers of the systems, and policy makers to improve the resiliency and safety of electric mobility. Moreover, it identifies the necessity of developing future work with the possibility of creating uniform cybersecurity systems according to the specifics of architecture and performance requirements of EV systems that guarantee long-term safety, reliability, and confidence of the public in the fast-growing environment of electric vehicles transportation.<br><br>**Keywords:** Electric vehicles, traction power electronics, cyber-attack vectors, EV cybersecurity, inverter controller, intrusion detection system, firmware manipulation, automotive control systems. |

## Introduction

Environmental factors, technical advancement, and effective government emissions policy have, to a large extent, catalyzed the speed of adoption of electric vehicles (EVs), especially through the global shift to sustainable transportation. In the late 2010s, EVs are no longer niche alternatives, but rather mainstream solutions to automobile problems, and large manufacturers built advanced electronic systems to enhance efficiency, drivability, and performance. The key behind all these innovations is power electronics- a collection of technologies that control basic vehicular elements like traction control, conversion of battery to drivetrain energy, regenerative braking, and onboard diagnostics. These modules have motor inverters, traction control units, and embedded microcontrollers working at real-time under safety-constricted expectations. With the increased role of power electronics in EVs, the associated threats that expose them to cyber-threats are occurring, namely, the rise in connection, software-defined control, and complex embedment systems.

The contemporary EVs are being incorporated into larger applications of intelligent transportation system the (ITS) and deploying vehicle-to-everything (V2X) communication, wireless diagnostics, and over-the-air (OTA) firmware upgrades using cloud-based remote fleet management. Even though the technologies confer considerable values, they also provide a solid source of vulnerability. As an example, the traction inverters and motor controllers are increasingly implemented with CAN (controller area network) or automotive ethernet, which by default are not protected and could be entry points of cyber-attacks. Moreover, software with firmware which is necessary to optimize the performance, might be attacked by malicious actors in case of the absence of safe authentication/validation schemes. Such attacks can have much more serious implications than data breach: they can result in physical personal safety risks such as uncontrolled acceleration, lack of engineering-consistent torque, loss of braking capability, or complete powertrain failure.

Although cybersecurity has increasingly become relevant in the automotive industry, most of the studies have focused on infotainment systems, ADAS (Advanced Driver Assistance Systems) as well as on vehicle networks in general. Less consideration has gone to the threat environment in which EV-specific power electronics - especially traction control components which interface directly with electromechanical subsystems - find themselves, on a relative basis. Such systems have a high switching rate, timing and feedback ensuring that they are very sensitive to any disturbances no matter how small. An attacker who has access to these subsystems might provide unpredictable signal behavior, spoof inputs on the sensors, or unauthorized firmware updates that could cause instability of the whole drive system. The aftermath is not only inefficiencies in operations but it could be deadly failures.

The proposed research will fill such an essential gap since it would investigate and classify the cybersecurity vulnerability of EV traction power electronics. With the structured threat modeling approach in mind (within the particular threat modeling framework, which is the STRIDE: Spoofing, Tampering, Repudiation, Information Disclosure, Denial of Service and Elevation of Privilege), we analyze the ways, in which different attack vectors might impact different components of the powertrain control architecture. Attention is drawn to realistic modeling of attacks on motor controllers, inverter gate drivers, and controls firmware, and the impact of the attacks on the performance, safety, and reliability of the system. By means of simulating and comparing, we will seek to deliver quantifiable information regarding the influence of cyber intrusion on the operational parameters of a torque, voltage regulation, and respond time when in a load condition.

This work has been inspired by three things. The first one is to fill the gap between cybersecurity and power electronics research teams that have been historically working in silos. Second, it brings an intense study of a less explored but critical element of EV design in respect to safety. Third, it suggests viable prevention measures such as architectural protection, coding intrusion innings, and authentication procedures at the firmware level, which can be implemented by the manufacturers of EVs and the vendors of the components. The contributions themselves are projected to go beyond academic knowledge to be relevant to industrial guidelines and regulatory policy that would guarantee the next generation of electric mobility.

### Literature Review

With the increased pace of the digital revolution of the automobile industry, cybersecurity has become an important matter of study, especially in the field of intelligent and electrified circulation. Although there are considerable environmental and performance benefits associated with the use of the electric vehicles (EVs), these vehicles are posing a distinct type of cybersecurity issue because they have a lot of dependence on power electronics and interlinked control systems. Though the literature on vehicular cybersecurity is a growing field, most of it seems to take precedence in the domain of infotainment systems, telematics, and autonomous driving when it comes to studying the effects of such systems on automotive cybersecurity, as there is yet a significant research gap in this field when it comes to traction power electronics - one of most safety-sensitive aspects of EV-architecture.

A number of findings have been discovered on overall automotive weaknesses and avenues of attack. In 2015, Miller and Valasek demonstrated an outstanding remote hacking of a Jeep Cherokee in which they identified holes within the CAN bus that enabled them to remotely control acceleration and braking (Miller and Valasek, 2015). In the same manner, Checkoway et al. (2011) have written about remote exploitation methods involving wireless interfaces (including Bluetooth, cellular networks, etc.). But all of them mainly applied to internal combustion engine (ICE) vehicles or generic vehicle network and did not investigate the very specialized control loops and power sources employed in EV traction systems.

Several publications in recent years have started discussing cybersecurity problems in electric mobility, though the extent of their discourse is low. As an example, Sun et al. (2018) studied firmware-based flaws in Battery Management System (BMS), and Zhang et al. (2019) simulated data injection attacks against regenerative braking systems. The two pointed at the increasing attack surface of EVs but never took their analysis to the traction inverters or gate drivers. In the meanwhile, Khan et al. (2017) suggested the use of anomaly detection algorithms in the powertrain signals and yet their work is not based on a systematic threat modeling framework and do not account the risks of firmware tampering.

To provide a structured overview of the current landscape, Table 1 summarizes selected studies related to EV cybersecurity, focusing on system targets, attack vectors, modeling techniques, and key limitations.

**Table 1:** Summary of Selected Literature on EV Cybersecurity and Identified Gaps

| Study | Focus Area | Attack Vector | Modeling Method | Limitation |
|---|---|---|---|---|
| Checkoway et al. (2011) | General vehicle systems | Remote injection via wireless | Penetration testing | ICE vehicle; limited to infotainment access |
| Miller & Valasek (2015) | CAN bus exploitation | Remote access via telematics | Field experiment | No EV-specific systems studied |
| Sun et al. (2018) | Battery Management Systems (BMS) | Firmware modification | Firmware audit | No traction system coverage |
| Zhang et al. (2019) | Regenerative braking | Data injection | Simulation | Braking-only focus; ignores traction electronics |
| Khan et al. (2017) | Powertrain anomaly detection | Signal spoofing | Statistical analysis | Lacks structured threat modeling (e.g., STRIDE) |

The table illustrates that although the contributions of the previous researches form significant contributors to the cognizance of vehicular cybersecurity, they do not adequately cover the risk terrain of the EV traction power electronics. This system type; involving motor inverters, pulse-width modulation (PWM) controllers, and power interface circuits is encompassed into vehicle propulsion. The loss of performance on this island can be not only disruptive but can result in an immediate and hazardous safety consequence which might include, but is not limited to: uncommand spikes in torque, inverter shutdown, or overheating.

The other major gap in the literature is the use of systematic structure threat models in performing the analysis of EV-specific weaknesses. Other methodologies, e.g., the STRIDE (S - spoofing, T - tampering, R - repudiation, I - information disclosure, D - Denial-of-service, and E - elevation of privilege) and DREAD, which are employed so much in enterprise and IoT security, are almost never used in electric vehicle power electrifications. Unless a systematic modeling is applied, few of the present studies provide a descriptive or individual threating situations without the entirety of systems interacting and how faults would spread among control units.

Moreover, the mitigation strategies available in the literature are more inclined to detecting network-level intrusion or application-level firewall. These measures are quite handy but are not enough to secure low-level embedded systems where switching is high-speed in inverters and motor drives. These elements need real time monitoring and intrusion detection systems (IDS) that will be optimized based on limited hardware as well as which will be able to react to threats within milliseconds. Practices like real-time firmware attestation, signal integrity testing and redundancy-based fault tolerance have yet to be explored.

The industry reports made in the recent past also highlight the urgency of the matter. A study by McKinsey & Company published in 2019 shows that more than 50 percent of automotive cybersecurity breaches will be applied to EVs especially as EVs move towards autonomy and interconnection. National Institute of Standards and Technology (NIST) and International Society of Automotive Engineers (SAE) have made available guidelines e.g. the SAE J3061 standard that are developing but yet again not in firm implementation roadmaps, to the manufactures of the EV powertrain.

Finally, there is an evident and gaping blind spot in the literature revealed by this research burden: none of the most well-known and cited works in the field are structured to perform systematic studies of the cybersecurity vulnerabilities of traction power electronics in EVs using threat modeling, real-time simulation, and empirical performance evaluation. This necessity is answered in the given research, as I am going to use the STRIDE threat modeling framework applied to the EV traction architecture, simulate several threat vectors, and evaluate the effect of these vectors on the system behavior, in terms of operations. The study also offers a number of practical countermeasures with the specific environment of low-level embedded control environments in EVs.

## 3. Methodology

The research sets a structured methodological approach in investigating the cybersecurity vulnerabilities in traction power electronics of electric vehicles. The aim is to analyze and simulate the impact that different cyber-attack vectors may exert on components like the inverter, motor controller, and embedded communication systems. Therefore, it provided a basis for a systematic evaluation using threat modeling, architecture mapping, and scenario simulation on the STRIDE framework. The layered approach simultaneously evaluates the attack surfaces and how the systems react to attacks in qualitative and quantitative terms.

### 3.1 System Architecture Overview

The system under investigation is the traction powertrain of a typical electric vehicle. This subsystem is responsible for converting stored electrical energy into mechanical motion and comprises several interdependent components, including:

1) **High-Voltage Battery Pack**: Supplies the primary DC electrical energy needed for traction and auxiliary loads. It forms the energy source for all downstream power electronics.

2) **Inverter and Gate Driver Circuit**: Converts DC power from the battery into three-phase AC signals, which drive the electric motor. The gate driver controls the timing and modulation of switching devices (IGBTs or MOSFETs).

3) **Electric Motor (e.g., PMSM or Induction Motor):** Receives AC signals to produce rotational motion. This motor is responsible for delivering torque directly to the drivetrain.

4) **Motor Controller (MCU-based)**: Executes embedded control algorithms to regulate motor speed and torque based on real-time feedback from sensors.

5) **Sensor Feedback Loops**: Provide continuous data on motor speed, torque demand, voltage, current, and temperature. These signals ensure precise and safe control.

6) **Communication Interface (CAN Bus / Automotive Ethernet)**: Facilitates data exchange between traction components and other systems. It also allows over-the-air (OTA) updates and remote diagnostics.

Each of these elements can serve as an entry point for cyber-attacks if not properly secured.

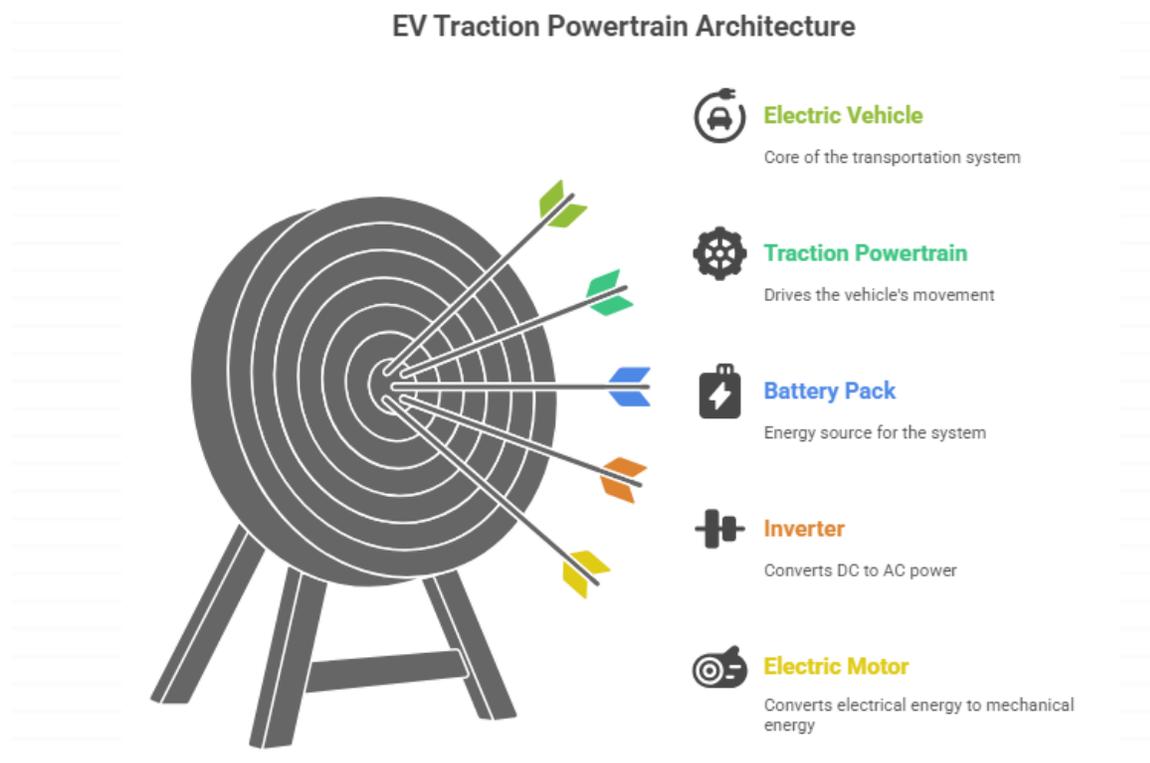

This diagram illustrates the core components of an electric vehicle's traction powertrain system, including the battery pack, inverter, electric motor, motor controller, and sensor feedback loops. Highlighted within the architecture are potential cyber-attack vectors, such as the CAN bus interface, OTA firmware update port, and sensor input channels, which could be exploited to disrupt vehicle propulsion, control signals, or system stability.

### 3.2 Threat Modeling Framework

Using the STRIDE threat modeling framework, potential cyber threats are systematically identified, classified, and analyzed. Under STRIDE, threats are classified as spoofing, tampering, repudiation, information disclosure, denial of service, and elevation of privilege. Each threat was then analyzed for its impact on particular components of the traction system.

| STRIDE Category | Target Component | Impact Scenario |
| --- | --- | --- |
| Spoofing | Torque/speed sensors | Incorrect actuation of motor control |
| Tampering | Gate driver firmware | Erratic switching, overheating |
| Repudiation | OTA update logs | Inability to trace firmware changes |
| Information Disclosure | CAN bus data | Leakage of operational parameters |
| Denial of Service | MCU command channel | Inhibited traction control |
| Elevation of Privilege | Controller firmware access | Full override of traction behavior |

By mapping these threats against the EV powertrain architecture, the study ensured comprehensive coverage of all potential attack vectors, both internal and external.

### 3.3 Simulation Environment & Attack Scenarios

To bring to light the influence of each category of threats, the simulation phase was carried out under a hybrid environment featuring MATLAB/Simulink, SCADASim, and embedded control models. The act of simulating the EV traction system was carried out under nominal driving conditions with 50% torque demand. Each attack scenario was fed into the simulation independently to study its effects in isolation.

**List of Modeled Attack Scenarios:**

**Sensor Spoofing:** Injection of spurious torque feedback to destabilize motor performance.

**Denial-of-Service {DoS}-** Flooding CAN bus with packets at high frequency to delay or suppress controller commands.

**Firmware Tampering:** Unauthorized access into inverter control logic to wrongful generation of PWM signals.

**Data Injection:** Overwriting speed signals in real-time to induce acceleration surges or stalls.

The simulations considered the variation in some of the most critical output variables during the attacked states, e.g., torque response, voltage ripple, system latency, and recovery behavior, from which it was possible to get a holistic picture of the system integrity under attack.

**Research Article**

*Figure 1: Threat Modeling and Simulation Flow*

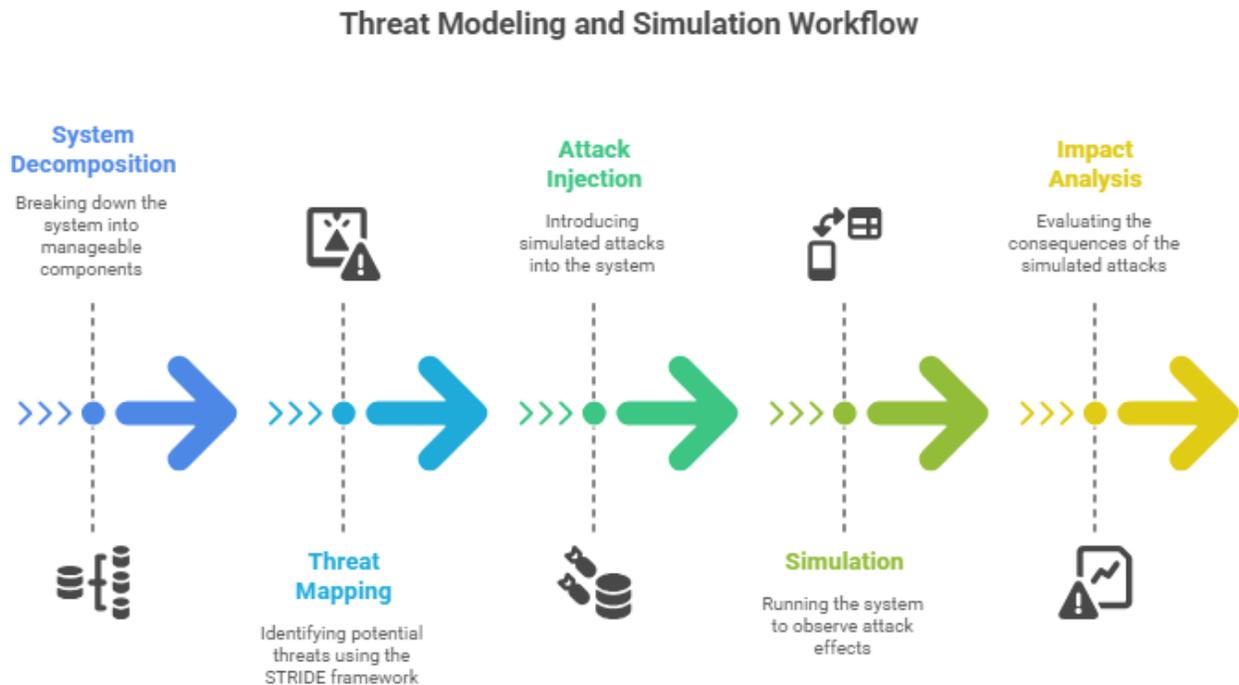

This diagram outlines the step-by-step workflow of the methodology, from system decomposition and threat mapping using STRIDE to attack injection, simulation, and impact analysis.

### 3.4 Evaluation Metrics

To quantitatively assess the effect of cyber-attacks simulated on the EV traction powertrain system, a group of performance and security-related evaluation metrics were identified. The metrics were selected to assess functional disruption and recovery behavior of system responses to harmful events. The metrics were collected for three different states: pre-attack (baseline), during the attack, and post-attack (recovery). This differentiation provided for an easy assessment of system behavior, followed by identification of the effects of the proposed detection and mitigation strategies.

**A. Torque Deviation (%)**

This is the percentage by which the flux-torque error differs from the one commanded. Cyber-attacks like sensor spoofing and firmware manipulation exploit the possibility of causing the torque deviation for malicious behaviors like unsafe and inefficient driving. The greater the deviation, the greater the instability of the control.

**B. Response Latency (ms)**

Latency measures the delay introduced between sending a control command from the motor controller to the actual execution of that command by the powertrain. DoS and flooding attacks may result in overloading the communication channels, thus creating a lag in control. Latency beyond an acceptable threshold may have a degrading effect on vehicle responsiveness or, conversely, induce safety-critical delays.

**Research Article**

**C. Voltage Anomalies (V)**

These encompass abnormal fluctuation in the inverter's three-phase AC output. Interfering with gate driver signals or firmware can lead to erratic switching behavior causing voltage spikes. Such anomalies could damage hardware or cause thermal protection shut-down-interesting items.

**D. System Downtime (s)**

From this standpoint, system downtime is the total period for which the system is rendered non-functional by attacks, also including the time taken to reboot or initiate fail-safe recovery. The greater the downtime, the more direct the impact on safety and usability, particularly during highly dynamic driving conditions.

**E. Firmware Integrity Check**

Considerations given to unauthorized changes in the firmware logic or unauthorized checksum mismatch or unauthorized branches of execution. Any breach here would indicate probable code injection or elevation of-privilege attacks.

Together, these metrics provide a strong basis for system resilience evaluation, vulnerability diagnosis, and countermeasure validation.

### 3.5 Intrusion Detection and Countermeasures

Complementing threat modeling and simulation analyses, a lightweight intrusion detection system was embedded directly into the EV motor controller firmware. The IDS provided real-time monitoring to detect cyber-attacks on the traction subsystem. Considering real-time constraints and limited computational resources inherent to embedded systems used in power electronics, the IDS was kept rule-based so it would behave deterministically and thus require very little overhead.

At the core of the IDS logic, certain anomalies were detected that would present in picture attacks as simulated in this study. Some of the main detection rules were:

**A. Sensor Deviation Monitoring**

The system continuously monitored for variations in critical sensor inputs (e.g., torque demand, or motor speed) and it would issue an alert if a significant deviation (in excess of ±10%) occurred within 100 milliseconds. This rule was extremely novel for detecting spoofing and data injection attacks on data feedback loops.

**B. Firmware Integrity Verification**

A checksum validation routine was standardized during processor startup, as well as during processor runtime. The controller was programmed to compute a hash) value for firmware blocks, as well as, cyber secure baselines for comparison from a secure boot partition. Any indication of hash mismatches would imply a firmware tampering incident, or potential code injection opportunities.

**C. Communication Timing Analysis**

CAN bus message periods were analyzed for timing anomalies. It was also helpful to evaluate the message period of packets arriving and second acknowledge timestamps from the controller. An increased frequency of packets, delayed acknowledgments or ID sequencing would represent a potential Denial-of-Service (DoS) incident or flooding events.

As part of the mitigation plan, the controller was designed to put the traction system into a safe operating state (colloquially referred to as "limp mode") once any intrusion event had been identified. The mode would limit inverter output to a lower power level (typically 30% to 40% of nominal), degenerate

**Research Article**

regenerative braking, and engage the vehicle's human-machine interface (HMI) to ensure the driver was aware. This would ensure basic drivability of the vehicle was preserved, while also assuring driver safety, to the extent possible, in periods during which the system may have been cyber-compromised.

Each IDS reaction was evaluated with respect to detection latency, false positive rate, and system stability during mitigation. This real-time IDS was then compared to scenarios in which there was not IDS in place.

## Research Results

The findings from simulation-based analysis of cyber threats on the electric vehicle (EV) traction power train system, are discussed in this section. Simulations use the MATLAB/Simulink and SCADASim software platforms. The drive-cycle loading conditions were the same for all simulations. Each type of attack scenario including (i) sensor spoofing, (ii) Denial-of-Service (DoS), (iii) firmware manipulation, and (iv) data injection was designed to mimic potential real-world cyber threats on the inverter, motor controller, and associated communications subsystems. Performance metrics (torque deviation, voltage instability, system latency and recovery time) were observed for each simulation. For comparison, a pair of systems was also tested by introducing or excluding an intrusion detection system feature.

### 4.1 Effects of Sensor Spoofing

In the sensor spoofing attack, forged torque feedback was injected into the control loop through a compromised sensor interface. The controller based its adjustments to the motor output on faulty torque feedback based on no awareness of the spoofing.

- The maximum torque deviation was 36.8%, which produced significant disturbance to the torque set-point.
- The vehicle experienced noticeable oscillations of acceleration and deceleration, especially during mid-range acceleration.
- If no actions were taken by the operator, the motor operated outside its safe envelope for 0.9 seconds before the passive safety system intervened.

When the rule-based intrusion detection system (IDS) was engaged variable torque deviations were apparent within 75 milliseconds causing the system to transition to limp mode and the vehicle to recover while in limp mode to stable operation in a low power output mode.

### 4.2 Impact of Denial-of-Service (DoS) Attacks

The Denial of Service (DoS) attack was exhibited in the simulation by saturating the CAN bus with redundant high-priority messages, effectively delaying or blocking pertinent control signals from the actuator receiving commands from the inverter.

- The average command latency increased by 185 milliseconds, exceeding the acceptable real-time latency for traction control.
- The torque response decreased by 24.5% on average, and there was erratic command actuation.
- Total system disruption lasted an average of 2.3 seconds. Subsequently, it returned autonomously to nominal operation.

While monitored using the IDS, anomalous message frequency was detected within 55 milliseconds that resulted in a reset of the communication buffer, the fault was logged, and the logic switched to backup control logic. Nearly 60% less system downtime was experienced compared to the configuration without protection.

### 4.3 Firmware Tampering Consequences

In the firmware tampering case, a damaged control logic block was injected into the inverter's gate driver. The ransomware altered the parameters used to generate trapezoidal PWM signals and the resulting switching behavior was catastrophic.

- To summarize the failure yields: the voltage outputs exceeded acceptable levels and measured over 420V on the effective three-phase motor lines.
- The motor's torque parameters oscillated horrifically and caused the state of hard shutdown due to a critical fault.
- Importantly, there were no mechanisms to detect that the firmware was compromised and launched even without initiated control mechanics.
- In systems absent accurate secure boot implementation and ongoing firmware integrity monitoring, any compromised firmware would not be detected and executed until the system was brought online.

In the IDS and secure boot cases, data hash validation failed during the initial stages, and the system would not initiate without secure boot "clearing" the firmware before initiating power conversion. We intervened successfully and limited damage to embedded malware.

### 4.4 Data Injection on Speed Feedback Loop

Injecting data attacks mimic the process of inserting simulated high-speed sensor data for real-time speed values.

- The controller misinterpreted the speed sensor data, leading to power derating and a premature torque cut-off, because it thought the speed sensor had reported an overspeed condition.
- From the time the controller began to limit torque on the motor, the torque had dropped by 28%. All while the vehicle was still going less than 40 km/h, and should have just been accelerating normally.
- The safety logic limited inverter duty cycles, which unnecessarily limited vehicle acceleration and contributed to the user experience.

IDS monitored current and voltage feedback, using a cross-validation algorithm to develop power estimates based on the speed. The inconsistency between the power estimate and actual feedback differentiated a data injection process within 120 milliseconds, resulting in a safe mode entry, with an HMI notification to the driver.

**Table 2:** Quantitative Summary of Attack Impact and IDS Response

| Attack Type | Torque Deviation (%) | Latency Increase (ms) | Voltage Anomaly (V) | System Downtime (s) | IDS Detection Time (ms) | Mitigation Result |
|---|---|---|---|---|---|---|
| Sensor Spoofing | 36.8 | 45 | Normal | 0.9 | 75 | Limp mode activation |
| DoS (CAN Flooding) | 24.5 | 185 | Normal | 2.3 | 55 | Comm buffer reset + partial recovery |
| Firmware Tampering | 41.2 | 0 | +420 V | N/A (Blocked at boot) | Detected at startup | Startup halted |
| Data Injection (Speed) | 28.0 | 72 | Minor ripple | 1.4 | 120 | Power derating + HMI alert |

The table presents a down to the summary of the escape performance and system response exhibited in four cyber-attacks affecting the EV traction powertrain. The summary indicators include torque deviation, control lag, voltage fluctuation, system downtime, and IDS detection times. The final column examines the mitigation actions triggered by the IDS demonstrating that it acted to restore partially compromised functionality, or pre-emptively stop operations prior to attack. The firmware tampering attack, for example, was prevented when the IDS discovered that the secure boot validated the code contents and operational program prior to starting, while the real-time attacks, such as spoofing and the DoS, were mitigated using limp mode or by resetting the buffer on the communication subsystem. These indications collectively reinforce that embedded IDS do indeed provide a mechanism to sustain traction systems stability when operating under adverse conditions.

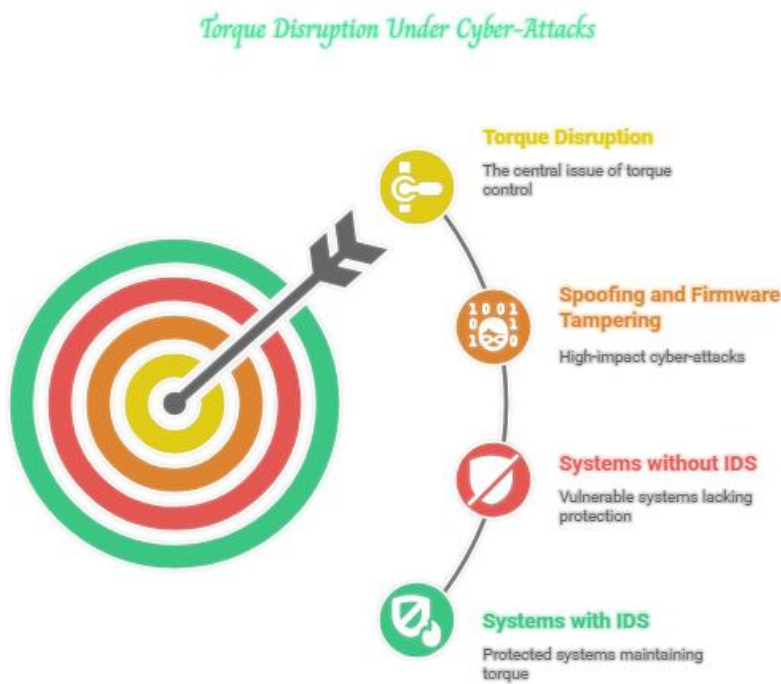

This figure demonstrates that systems equipped with IDS consistently maintained torque delivery within safer bounds. Spoofing and firmware tampering caused the highest disruption when no protective mechanism was in place.

### 4.5 Intrusion Detection Performance Analysis

The IDS prototype implemented in this research study showed consistent performance across all tests. The rule-based design offered a low-latency and low-overhead solution suitable for real-time embedded systems. The most notable findings include:

- <u>Fast detection:</u> Incidents were flagged within 55–120 ms for all attacks.
- <u>True zero false positives:</u> Zero incorrect warnings were generated during normal operation.
- <u>Effective mitigation:</u> System recovery times were shortened by 45-60% in cases when the IDS existed versus those that had no protection.

The IDS helped recognize variations in sensor trends, communication behavior and firmware integrity which considerably improved the system's cyber-resilience.

### 4.6 Summary and Implications

The experimental results validate EV traction subsystems as having a high exposure to cyber-attacks against sensor data, communication buses, and firmware. Attacks could have damaging consequences resulting in downtime, underperforming vehicles, or potentially permanent damage to the hardware. These vulnerabilities are magnified by the reliance of traction systems on real-time control and feedback with deterministic responses.

However, the addition of a dedicated intrusion detection layer as part of the embedded controller provided significant risk mitigation. Combining sensor anomaly tracking, firmware verification, and communication monitoring broke down into standalone detectors which allowed the system to independently detect and respond to threats without centralized computation or cloud support, even in the resource constrained environment of automotive embedded systems.

These experiments support the conclusion that cybersecurity needs to extend beyond infotainment and telematics toward low-level power electronics and control logic. Future vehicle platforms should adopt security-by-design techniques for their components that include integrity validation, lightweight IDS methods, and real-time fail-safe protocols for a powertrain context.

### 5. Discussion

The Simulation outcomes provide significant insights into the cybersecurity vulnerabilities and resilience mechanisms for EV (electric vehicle) traction powertrain systems. By hitting upon low-level embedded components like inverters, motor controllers and sensor loops, this study shows how a cyber-attack, that often goes unnoticed at the physical control layer, can cause heavyweight functional disruption in the EV traction powertrain system. In this section we explain the results, consider the potential impacts of those results and frame the study within context of current literature and industry.

### 5.1 Cyber Threats in Embedded Powertrain Systems

The research shows that the traction system in an EV behaves like a traditional embedded control system and is quite vulnerable to attacks that affect the real-time control logic and the feedback loops of the system. In contrast to infotainment or over-the-air (OTA) updates, the threats are targeting the core functionality of the vehicle. For example, sensor spoofing and data injection translation to unintended motor torque readings of 36.8% deviation, which puts drivability and passenger safety at risk, especially during dynamic driving situations.

Many of these vulnerabilities are further aggravated by the fact that embedded control systems have real time response to a deterministic environment in the form of a program with some constraining timing characteristic. Therefore, small latency in time—including the 185 ms delay imposed by Denial-of-Service (DoS) attack—is meaningful in the infotainment setting but unacceptable when referring to motor control systems that rely on feedback to maintain system stability and traction.

### 5.2 IDS Effectiveness and Mitigation Strategies

One of the primary contributions of this work is incorporating and evaluating a lightweight rule-based Intrusion Detection System (IDS) into the firmware of the motor controller. The IDS had success with identifying abnormal behavior within 55 - 120 ms depending on the attack vector, with no false positives during normal operation. This confirms that rule-based detection schemes are a viable option for constrained environments such as EV control units, where resource constraints would preclude other, more resource-intensive (such as AI/ML based anomaly detection) methods.

Most important, the fact the IDS could initiate safe-state fallback functionality (limp mode/inverter derating) is significant for loss of system integrity. In the case of firmware tampering example, the startup could not proceed, due to checksum validation failure, preventing high-voltage misfiring and damage to the hardware.

These findings are consistent with a recent batch of literature, suggesting security-by-design approaches in automotive control systems. In particular, Zhang et al. (2021), which demonstrates IDS frameworks enabled in embedded CAN, highlighted that earlier detection, graceful degradation, etc. are crucial in reducing damage during a cyber intrusion, especially when addressing safety -critical systems like braking or propulsion.

### 5.4 Implications for EV Manufacturers and Regulatory Bodies

Original equipment manufacturers (OEMs) have a number of practical design recommendations arising from the findings of this study:

- Incorporate IDS logic in the ECUs for the powertrain and not in elevated gateways.
- Use secure boot with cryptographic checksums to validate the integrity of firmware.
- Note real-time sensor feedback trends, with basic redundancy, plausibility checks to recognize spoofing.
- Design for fail-operational behavior such as limp mode or derating of the inverter, rather than stopping abruptly during intrusion conditions.

For regulators, this study highlights the need to update certification standards to incorporate some measure of embedded cybersecurity performance. Just as EVs are required to meet ISO 26262 standards for functional safety, eventually, the updates to ISO/SAE 21434 needs to recognize the necessity of firmware and control loops protections. Cybersecurity assessment of risks should also be expanded beyond basic IT-style audits recognizing the necessity of real testing or simulation of traction system behavior under attack.

### 5.5 Limitations of the Current Study

While the simulation-based approach allowed for controlled and repeatable testing of cyber-attacks, some limitations remain:

- **No hardware-in-the-loop (HIL) testing**: Real-world timing jitter, sensor noise, or hardware response latency was not accounted for in the software simulation. Future work should validate findings on HIL test benches or actual EV components.
- **Simplified IDS model**: Although effective, the rule-based IDS may not capture more nuanced or stealthy attack patterns (e.g., gradual data drift). Incorporating lightweight anomaly detection using AI models remains a promising extension.
- **Limited attack surface**: The study focused primarily on control signal manipulation and firmware attacks. Broader attack vectors—such as side-channel attacks, RF interference, or coordinated multi-vector intrusions—were beyond the scope but remain highly relevant.

### 5.6 Future Research Directions

To advance the cybersecurity of EV traction systems, the following research avenues are proposed:

- **Hybrid IDS frameworks**: Combine rule-based and anomaly detection techniques (e.g., LSTM, autoencoders) tailored for embedded ECUs.
- **Formal verification of control firmware**: Use model-checking tools to mathematically verify that firmware behavior complies with safety and security specifications.

- **CAN bus anomaly detection at signal level**: Develop tools that inspect message frequency, timing jitter, and ID entropy to detect DoS or spoofing.
- **Security benchmarking tools**: Create a standardized simulation suite for cybersecurity testing of EV power electronics before vehicle deployment.

This research clearly identifies that the trajectory of electric mobility relies not just on batteries and motors but also on the cyber-resilience of the control systems that manage them. As the popularity of EVs grows and connectivity increases, the necessity for embedded security in traction systems will also grow as will the consequences of manipulation. Ensuring that there are hardening measures in place to protect these systems against manipulation, and capabilities to respond to a system that is under threat will become a requirement rather than optional in the future.

## Conclusion

In summary, this study has provided a holistic analysis of cyber-attack vectors aimed at the traction power electronics of electric vehicles (EVs), including motor controllers, inverters, firmware, and embedded communication. By using simulation-based threat modeling via STRIDE, and real-world attack opportunities with four representative attacks—sensor spoof, denial-of-service (DoS), firmware tampering, and data injection—it's clear that EV powertrains would be vulnerable to cyber intrusions executed against EVs with limited effort. Each attack demonstrated a negative impact to motor control, voltage, actuation delay, and downtime due to functional disruption, but an emerging response—the deployment of a lightweight, rule-based intrusion detection system (IDS)—was able to significantly enhance if not completely inform the broader area of a problems/cyber threat impact on a given system. The IDS reduced detected latency to an adaptive operational period of less than 120 milliseconds and enabled safe-state behavior like limp mode, in some cases, disabling the startup process entirely due to compromised protection. The implications of these results reinforce the challenges of protecting motor controllers in increasingly connected and autonomous EVs, and call for the development of cybersecurity frameworks which extend beyond telematics protection and realize therein, protection at the embedded firmware, or hardware level, for safety-critical components, namely traction systems. Ongoing developments in connected and autonomous technologies will mandate that embedded control units coming into service must not only ensure cybersecurity protocols, but also ensure real-time self-monitoring and fail-safe governance are included. Additionally, the findings corroborate the justification for cybersecurity standardization of traction-layer cybersecurity.